\documentstyle[pre,aps,epsf]{revtex} 

\begin{document}
\draft

\title{Nonequilibrium Phase Transition and Self-Organized Criticality\\ 
in a Sandpile Model with Stochastic Dynamics }

\author{S. L\"ubeck$^1$\cite{SvenEmail}, B. Tadi\'c$^2$\cite{BosaEmail} 
and K.~D. Usadel$^1$\cite{UsadelEmail}}

\address{
$^1$Theoretische Tieftemperaturphysik, 
Gerhard-Mercator-Universit\"at Duisburg,\\ 
Lotharstr. 1, 47048 Duisburg, Germany \\
$^2$Jo\v{z}ef Stefan Institute, University of Ljubljana,\\ 
P.O. Box 100, 61111-Ljubljana, Slovenia \\}

\date{14. September 1995}

\maketitle

\begin{abstract}
\newline
We introduce and study numerically 
a directed two-dimensional sandpile automaton with probabilistic 
toppling (probability parameter~$p$)  which
provides a good laboratory to study both self-organized criticality and
the far-from-equilibrium phase transition. 
In the limit $p=1$ our model 
reduces to the critical height model in which the self-organized 
critical behavior was found by exact solution [D. Dhar and R. Ramaswamy, 
Phys. Rev. Lett. {\bf 63}, 1659 (1989)]. 
For $0< p < 1$ metastable columns of sand may be
formed, which are relaxed when one of the local slopes exceeds a critical
value $\sigma _c$.
By varying  the probability of toppling  $p$ we find that a continuous 
phase transition occurs at the critical probability $p_c$,
at which the steady states with zero average slope (above~$p_c$) are 
replaced by  states  characterized by a  finite  average 
slope (below~$p_c$). 
We study this phase transition 
in detail by introducing an appropriate order parameter 
and the order-parameter susceptibility $\chi $. 
In a certain range of $p<1$ we find the self-organized 
critical behavior  which is 
characterized by nonuniversal $p-$dependent scaling exponents
for  the probability distributions of size and length of 
avalanches.
We also calculate the anisotropy exponent $\zeta $ and the fractal 
dimension  $d_f$ of 
relaxation clusters in the entire range of values of the 
toppling parameter $p$. 
We show that the relaxation clusters 
in our model are anisotropic and can be described as fractals for
values of $p$ above the transition point.
Below the transition they are isotropic 
and compact.
\end{abstract}
\vspace{-12cm}
{\tiny Phys.~Rev.~E {\bf 53}, 2182 (1996) }
\vspace{12cm}


\twocolumn

\section{Introduction}

The term "self-organized criticality" (SOC) \cite{BAK} refers
to certain driven dissipative systems which organize themselves
into a steady state far away from equilibrium. 
The critical steady state appears as  an attractor  of the dynamics,
and thus no  fine tuning of an  external parameter is necessary to 
achieve criticality.
The major aspect of the dynamics of SOC systems is the 
"separation of time scales", i.e. the system relaxes 
infinitely fast compared to the time between two successive perturbations.
From this point of view the relaxation processes may be
considered as consisting of a series of discrete events (avalanches)
and one investigates the probability distribution of their spatial 
properties, for instance the size of an avalanche.
The self-organized critical state   is
characterized by a power-law behavior of these probability
distributions.
For detailed discussions of the definition of SOC we refer to
\cite{GRINSTEIN,FLYVBJERG}.

Sandpile automata  \cite{BAK,DHAR,MANNA} 
are well known prototype models exhibiting SOC. 
By adding a particle from the outside, a sandpile is
perturbed and  the perturbation may lead to instabilities at neighboring
sites if the local height of sand exceeds a critical value $h_c$ (critical 
height model (CHM) \cite{BAK,DHAR}), or if the local slopes exceed a 
critical value $\sigma _c$ (critical slope model (CSM) 
\cite{MANNA,KADANOFF,LUEB_1}).
The steady state of the system in the case of the CHM is accompanied with
a state of a zero average slope. A finite average slope characterizes
a CSM.  

In Section II we introduce a two dimensional directed
sandpile model with stochastic dynamics, in which 
the  updating rules may be tuned continuously by varying 
a parameter~$p$, representing probability of toppling,
between CHM (for $p=1$) and CSM (for $p=0$) \cite{LUEB_2}. 
Our numerical results for $p<1$ show
that a nonequilibrium phase transition 
takes place at the critical value $p_c$, which is
characterized by a continuous appearance of a non zero average 
slope below the critical point. 
In order to characterize this phase transition
quantitatively, we introduce an order parameter 
$\langle \sigma \rangle$,
representing net average slope. 
Results obtained from numerical simulations of the order parameter 
$\langle \sigma \rangle$, as well as the order parameter 
susceptibility and the penetration depth of the slope will be
given in Section III.
 
We also study fractal properties of
the avalanches by calculating their fractal dimension 
in the entire range of values of the parameter $p$ (Section IV). 
Our results suggest that for values of $p$ above the transition point 
the avalanches can be described as self-affine fractals, i.e.
the shape of the avalanches exhibits an anisotropic behavior. 
Below the transition point both properties, the fractality and
anisotropy of the
relaxation clusters, are lost.

In Section V we present a detailed analysis of the probability
distributions of the avalanches and address the question
whether the model exhibits SOC for $p<1$.
A finite size scaling analysis of the probability distributions
and a check of certain scaling equations yielded
that the model displays SOC for certain values of $p<1$.

\section{Model}

We consider a two-dimensional sandpile model on a square lattice
of size $L \times L$ and integer variables $h(i,j)$ representing the local
height.
We assume a directed dynamics, i.e. particles are restricted
to flow in the downward direction (increasing $i$).
According to the widespread 'sandpile language', the first row $(i=1)$
and the last row $(i=L)$ represent the top and the bottom of the
pile, respectively.
To minimize the influence of the horizontal boundaries we limit our investigations
to periodic boundary conditions in this horizontal direction
($j$-direction).
Any site of the lattice has two downward and two upward next neighbours,
namely $h(i+1,j_{\pm})$ and $h(i-1,j_{ \pm})$ with 
$j_{\pm}=j \pm \frac{1}{2}(1 \pm (-1)^i)$.

We perturb the system by adding particles at a random place on the top
of the pile according to
\FL
\begin{equation}
h(1,j) \, \mapsto \, h(1,j)+1\, , \hspace{1cm} \mbox{with random }j.
\label{eq:perturbation}
\end{equation}
A site is called unstable if the height $h(i,j)$ or at least one of the 
two slopes 
\FL
\begin{equation}
\sigma (i,j_{\pm})=h(i,j)-h(i+1,j_{\pm})
\end{equation}
exceeds its respective critical value, i.e.
if $h(i,j) \ge h_c$ or $\sigma (i,j_{\pm}) \ge \sigma_c$.

If both slopes are critical one particle after another drops alternatively
to the next downward neighbours until both slopes become subcritical.
In the case that one slope exceeds $\sigma_c$ particles drop to the corresponding
downward neighbour.

In contrast to the critical slope, the critical height condition has
a stochastic character.
If the local height exceeds the critical value $h_c$ 
toppling occurs only with the probability $p$, and then two particles drop
to the two downward neighbours.

Because each relaxing cell changes the heights or  slopes 
of its four neighbouring sites, the stability conditions are applied at 
these four activated  sites in the next updating step.
Toppling may take place after adding a particle on the first row. 
We apply the slope and the height stability condition simultaneously
for all activated sites.
An avalanche stops if all activated sites are stable.
Then we start again according to Eq.~(\ref{eq:perturbation}).

It should be emphasized that, in addition to the instability criteria,
a site is considered as potentially unstable {\it only } when a
particle drops at or near the site. Owing to the probabilistic character
of the critical height toppling rule, locally heights could remain 
above critical after one relaxation event. In this respect, the present
model for $0<p<1$ is different from already known critical height 
\cite{BAK,DHAR,MANNA}
and critical slope  
\cite{MANNA,KADANOFF} models.

One can interpret the parameter $1-p$ as being 
due to  static friction between the sand grains, which prevents toppling 
even if the height exceeds the critical value.
In the case  $p=1$,  our model is identical to the model of Dhar and 
Ramaswamy \cite{DHAR}, which exhibits a robust SOC behavior.
In the  limit $p=0$,   our directed rules lead to a steady state where all 
slopes are equal, i.e. $\sigma(i,j_{\pm})=\sigma_{c}-1$. 
Thus, if a particle is added at the first row it  
performs a directed
random walk downward the pile until it reaches the boundary and
drops out of the system --- no SOC  can  occur in this limit.

We fix the critical height $h_{c}$
and the critical slope $\sigma_{c}$ 
and restrict ourselves to $h_{c}=2$ and $\sigma_{c}=8$,
although it should be emphasized that the results depend on
the choice of these parameters.

The behavior of our model is characterized
by anisotropy effects caused by the preferred
direction of the assumed dynamics. 
This anisotropy may influence for instance the fractal 
properties of the avalanches.
If a measured quantity depends on the direction we
introduce in these cases two different values which
describe the behavior in parallel and perpendicular
direction respectively, e.g. the fractal dimensions 
$d_\parallel$ and $d_\perp$.
The parallel direction corresponds to the preferred
direction of the dynamics (increasing $i$).

\section{Nonequilibrium Phase Transition}

In this section we consider the general behavior of the system, i.e.
how the transition from the CHM to the CSM takes place. 
As mentioned above a CHM is characterized by a zero average slope
in contrast to a CSM where the steady state displays a non zero
average slope.
In order to describe this transition quantitatively,
we introduce as an order parameter 
the  mean slope of the system \cite{LUEB_2}
\FL
\begin{equation} 
\langle \sigma \rangle = \frac{1}{L_{\parallel}L_{\bot}}
\sum_{i,j}   \left \langle  \sigma_{ij} \right \rangle
\label{eq:order}
\end{equation}     
where $ \langle \dots  \rangle $ stands for the average over
total number of events, and $\sigma _{ij}$ 
is the local slope at site $(i,j)$ which is defined as 
\FL
\begin{equation} 
\sigma_{ij}=\frac{1}{2}(2 h_{i,j}-h_{i+1,j_{+}}-h_{i+1,j_{-}}) .
\end{equation} 
The summation  in Eq.~(\ref{eq:order}) runs  from $j=1,2,3,...,L$
in perpendicular direction, 
and from $i=21,22...,L-20$ in parallel direction,
in order to suppress the boundary effects from 
the first and last row of the system. 
Of course the obtained results are independent of the value
of the cut off length, provided that it is large enough.
To normalize the order parameter
one has to choose $L_{\parallel}=L-40$ and $L_{\bot}=L$.
The  nature of the transition as well as the transition point 
$p_c$ should be determined from the $p$-dependence of the average
slope Eq.~(\ref{eq:order}) in the vicinity of $p_c$.

\begin{figure}
 \begin{minipage}[t]{7.5cm}
 \epsfxsize=7.5cm
 \epsfysize=7.5cm
 \epsffile{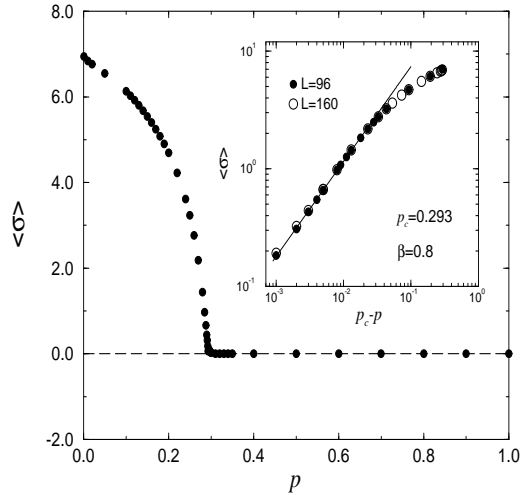} 
 \caption{Average slope of the pile $\langle \sigma \rangle$ plotted
	  vs. probability $p$. The solid line inside the inset 
	  corresponds to a fit according to Eq.~\ref{eq:order}. 
 \label{order_par}} 
 \end{minipage}
\end{figure}

In  Fig.~\ref{order_par} we show the results of numerical 
simulations of the order parameter defined in Eq.\ (\ref{eq:order}).
The order parameter vanishes at the transition point as, i.e.,
\FL
\begin{equation}
\langle \sigma \rangle \sim (p_c-p)^\beta ,
\label{eq:order_power}
\end{equation}
as one can see from the inset of Fig.~\ref{order_par}.
We determine the transition point $p_c=0.293\pm0.002$ and the 
exponent $\beta=0.8\pm0.05$ for automaton sizes
$L=64,96,128,160,192$ .

\begin{figure}
 \begin{minipage}[t]{7.5cm}
 \epsfxsize=7.5cm
 \epsfysize=7.5cm
 \epsffile{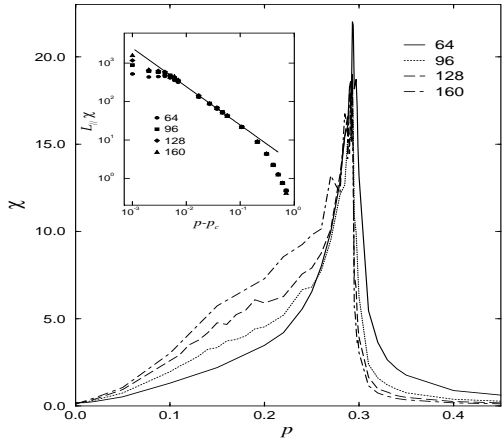} 
 \caption{The $p$-dependence of the order parameter susceptibility
	  for different system sizes $L$. 
	  The inset displays the scaling bahavior of the height regime.
	  The solid line corresponds to a power law with exponent 
	  $\gamma=0.98$.	  	  
 \label{suscept}} 
 \end{minipage}
\end{figure}

We next analyze the fluctuation of the order parameter. 
An ordering susceptibility can be defined as \cite{BINDER} 
\FL
\begin{equation}
\chi={L_{\parallel}L_{\bot}} \; \;
(\langle \sigma^2 \rangle - \langle \sigma \rangle^2). 
\label{eq:chi_binder}
\end{equation}
Unfortunately the average slope has as a self-averaging character 
in the critical height regime, that means $\chi$ vanishes for 
$L\rightarrow\infty$ due to a cancellation of the 
correlations  $\langle \sigma_{ij} \sigma_{kl} \rangle$ 
in the interior of the pile. 
Consequently the susceptibility depends
only on the boundary term and is of relative 
magnitude~$\sim L_{\parallel}^{-1}$. 
In the case $p=1$ the order parameter susceptibility
can be calculated exactly and is given by 
\FL
\begin{equation}
\chi=\frac{1}{2L_{\parallel}} 
\label{eq:chi_dhar}.
\end{equation}
In Fig.~\ref{suscept} we display the susceptibility $\chi$ for varying
$p$ and four different automata sizes. 
For all considered automata sizes
$\chi$ is sharply peaked at the transition point $p_c=0.293$. 

In the inset of Fig.~\ref{suscept} we plot $\chi \cdot L_{\parallel}$ vs. $p$.
All different curves collapse, except of the values which are close
to the transition point. The susceptibility scales 
as 
\FL
\begin{equation}
\chi \sim L_{\parallel}^{-1} (p-p_c)^{-\gamma}
\end{equation}
with $\gamma=0.98\pm0.05$.
One can see from Fig.~\ref{suscept} that this power law behavior of $\chi$
holds away from $p_c$ up to a certain cut off value
which tends to $p_c$ if $L\to\infty$.
Due to this $L$-dependence the order parameter susceptibility tends
to zero with increasing~$L$ for $p>p_c$.

Below the critical point our data suggest that the susceptibility 
displays a complicated scaling behavior (see Fig.~\ref{suscept}).
It is still an open question whether $\chi$ diverges with
increasing systems size $L$ or converges to
a finite i.e. $L$-independent value.
To address this question one has to simulate extremely large systems
which is beyond our present computer capacity.

In order to break off the self-averaging character of the average slopes
we defined a nonlinear-susceptibility
\FL
\begin{equation}
\chi_{_{nl}}={L_{\parallel}L_{\bot}} \; \;
(\langle \sigma_{_{nl}}^2 \rangle - \langle \sigma_{_{nl}} \rangle^2) 
\label{eq:nonlinear}
\end{equation}
with
\FL
\begin{equation} 
\sigma_{_{nl}} = \frac{1}{L_{\parallel}L_{\bot}}
 \sum_{i,j}  \sigma_{ij}^2 .
\label{eq:order_2}
\end{equation}  
We found that $\chi_{_{nl}}$ shows a cusp at $p_c$, i.e. the 
nonlinear-susceptibility is 
independent of $L$ in both regimes (not shown). 
This behavior is consistent with the exact value for $p=1$: 
\FL
\begin{equation} 
\chi_{_{nl}}=\frac{9}{64} + {\cal{O}}(L^{-1}). 
\label{eq:chi_dhar_2}
\end{equation}  

\begin{figure}
 \begin{minipage}[t]{7.5cm}
 \epsfxsize=7.5cm
 \epsfysize=7.5cm
 \epsffile{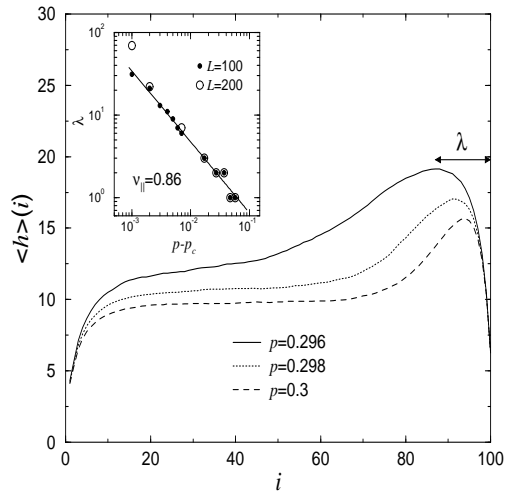} 
 \caption{The average height profile $\langle h(i) \rangle$ along the 
	  parallel direction. The penetration depth $\lambda$
	  is determined as the distance from the maximum height to the 
	  right boundary of the system. The inset displays the 
	  $p$-dependence of $\lambda$. 
 \label{pen_depth}}  
 \end{minipage}
\end{figure}

The non-diverging behavior of the susceptibilities indicate
that the non equilibrium phase transition is not characterized
by a diverging correlation length.
Thus we address the question how the finite slope 
occurs by approaching the transition point.
In Fig.~\ref{pen_depth} we display the average height profile 
$\langle h \rangle(i)$
of the pile for certain values of $p$. 
For $p<1$ a finite slope occurs at the boundaries and
the pile grows up from the bottom of the pile. 
The slopes penetrate the hole system from the edge of the
automaton.

We determine the penetration depth $\lambda$ of the slopes by measuring
the distance of the profile maximum from the right boundary
(see Fig.~\ref{pen_depth}).  
The obtained results are plotted in the inset of Fig.~\ref{pen_depth} for 
automaton sizes $L=100, 200$. 
Close to the transition point the penetration depth obeys
a power law behavior like
\FL
\begin{equation}
\lambda \sim (p-p_c)^{-\nu_\parallel} \; .
\label{eq:corr_par}
\end{equation}
Our numerical analysis yielded $\nu_\parallel=0.86\pm0.06$.

\section{Fractal Properties of the Avalanches}

In this section we consider the scaling properties of
the avalanches.
We define the size $s$ of the avalanches as the number
of toppled sites in one event.
The length of an avalanche is defined as the distance
from the first row to the toppled site with maximal value of
the index  $i$.
Note  that this length is by definition parallel
to  the preferred direction, and will be  denoted  as $l_\parallel$.
We introduce the average width $l_\bot$ of an avalanche
via
\FL
\begin{equation}
l_\bot = \left ( \; {l_\parallel}^{-1} \; \;
\sum_{i=1}^{l_\parallel} \;l_\bot(i)^q \; \right )^{\frac{1}{q}},
\label{eq:defwidth} 
\end{equation}  
where $l_\bot(i)$ is the width of an avalanche in the
$i$th row.
Since there is some arbitrariness in the definition of $l_\perp$
we both tried $q=1$ and $q=2$ with the result that 
the scaling properties of the width are practically the same
in both cases.
For the sake of simplicity we present in this paper only the results
for $q=2$.
We assume that the size of avalanches is scale invariant both 
with the length  and width.
In order to take the anisotropy of avalanches into account, we
introduce two different scaling exponents $d_{||}$ and $d_{\bot }$,
corresponding to different directions of scaling $l_{||}$ and $l_{\bot }$,
respectively \cite{KINZEL,KAISER}.
We assume that the average size  $\langle s \rangle$ of all 
avalanches of length $l_\parallel$ and of width $l_\bot$, 
respectively, scales as
\FL
\begin{equation}
\langle s \rangle _{l_\parallel} \sim  {l_\parallel}^{d_\parallel}
\label{eq:size_length}
\hspace{0.5cm} \mbox{and} \hspace{0.5cm}
\langle s \rangle _{l_\bot} \sim  {l_\bot}^{d_\bot} \;.
\label{eq:size_width}
\end{equation} 
If both scaling relations are valid the width scales with the 
length like
\FL
\begin{equation}
l_\bot \sim  {l_\parallel}^{\zeta}, \quad \mbox{with} \quad 
\zeta=\frac{d_\parallel}{d_\bot}. 
\label{eq:width_length}
\end{equation}  
where the exponent $\zeta$ describes the anisotropic behavior 
of the avalanches.
Since there are only two independent exponents we measured 
$d_\parallel$ and $\zeta$ according to Eq.~(\ref{eq:defwidth})
-Eq.~(\ref{eq:width_length}).

We find that with decreasing $p$ the parallel fractal dimension $d_{||}$ 
increases from the exactly known value $d_\parallel=\frac{3}{2}$ in the 
limit  $p=1$  to two at the transition point. 
In the critical slope regime
$d_{\parallel }$ remains  constant and $d_{\parallel}=2$ \cite{LUEB_2}.

\begin{figure} 
 \begin{minipage}[t]{7.5cm}
 \epsfxsize=7.5cm
 \epsfysize=7.5cm
 \epsffile{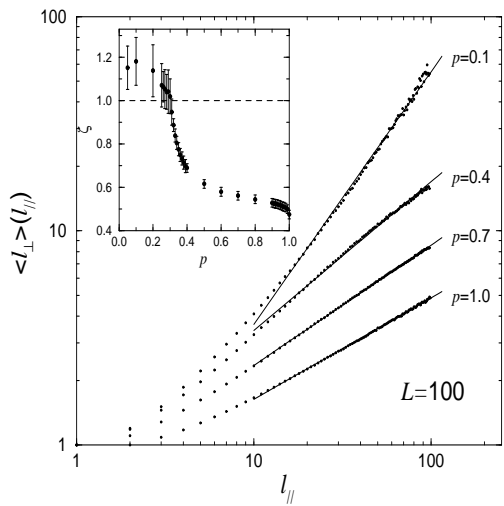} 
 \caption{Double logarithmic plot of the average width vs. length
	  for various values of $p$ as indicated. 
	  All curves obey a power law behavior (solid lines) with
	  different accuracy. 
	  The inset displays the $p$-dependence of the anisotropic 
          exponent $\zeta$ obtained from the fits according to 
	  Eq.~\ref{eq:width_length}.
 \label{width}}  
 \end{minipage}
\end{figure}

Fig.~\ref{width} shows how the width depends on the length for different 
values of $p$. 
In all cases we found that Eq.~(\ref{eq:width_length}) is valid and 
extracted values of the anisotropy exponent $\zeta$ (see inset to 
Fig.~\ref{width}). 
With decreasing $p$ the anisotropy exponent $\zeta$ increases 
from $\zeta=\frac{1}{2}$  at $p=1$ and reaches its maximum value 
$\zeta \simeq 1$ for $p \le p_c$.
We expect that this expression is exactly given by $\zeta=1$ and 
that the observed deviations in the critical
slope regime are caused by finite size effects.
The origin of the huge error bars for $p<p_c$ is a finite curvature
of the corresponding curves in the log-log plots.
These curvatures would 
be reduced by studying larger systems leading eventually to the result
$\zeta=1$ with  higher accuracy.
Thus,  the critical height regime is characterized by an
anisotropic shape of the avalanches $(\zeta < 1)$.
This behavior changes to an isotropic shape $(\zeta=1)$ on 
approaching the transition point.
Due to the dependence of the scaling factors on the directions
the avalanches can be regarded as an example of self-affine fractals.
In analogy  to self-similar fractals, it is possible to describe 
self-affine fractals by single exponent $d_f$, given by \cite{KAISER} 
\FL
\begin{equation}
d_f = 2 \frac{d_\parallel}{\zeta+1}.
\label{eq:fractal_dim}
\end{equation}

\begin{figure}
 \begin{minipage}[t]{7.5cm}
 \epsfxsize=7.5cm
 \epsfysize=7.5cm
 \epsffile{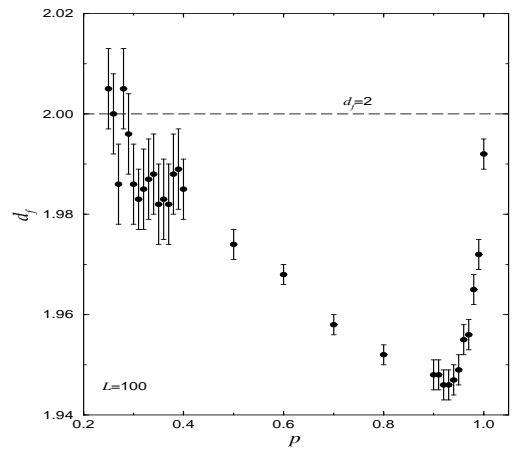} 
 \caption{Fractal dimension $d_f$ of the avalanches for various values
	  of $p$. 
 \label{frac_dim}}
 \end{minipage}
\end{figure}

In Fig.~\ref{frac_dim} we show the results for the fractal dimension $d_f$ 
versus $p$ in the critical height regime.
As seen from  Fig.~\ref{frac_dim}, the relaxation clusters exhibit a fractal 
behavior ($d_f<2$) in the critical height regime $p>p_c$, 
with  exception  of the case $p=1$. 
Although the deviation from $d_f=2$ is very small they cannot
be explained by statistical fluctuations due to the size
of the error bars and the systematic way of the deviations from $2$.
Below  the transition point, i.e., for
$p<p_c$, the fractal dimension $d_f =2$, corresponding to
compact clusters.

One can easily explain this behavior  by considering the
shapes of the avalanches (see Fig.~4 in \cite{LUEB_2}).
In the pure critical height model ($p=1$) the avalanches
are compact, i.e.~no holes can occur in this limit.
With decreasing $p$ some supercritical sites may remain
in the interior of the avalanche due to the stochastic
character of the dynamics, leading to holes and branching
processes. This results in a fractal character of the avalanches.
By approaching the transition point $p_c$ this structure
of the avalanches is lost.
In this region the dynamics is more and more characterized
by toppling processes due to the critical slope condition.
The slope between a toppled site and its two backward neighbours
may exceed the critical value and toppling occurs again.
Owing to this multiple topplings 
the probability to find holes inside an avalanche decreases 
with decreasing $p$ and eventually vanishes at $p_c$.
At the same time branching of the relaxation clusters become
more rare.

In this way,  we may conclude that the phase transition from the 
critical height to the critical slope regime is accompanied by 
continuous change 
from fractal avalanches with an anisotropic shape above $p_c$ to
compact and isotropic avalanches below $p_c$.

\section{Avalanches Distributions}

\begin{figure}
 \begin{minipage}[t]{7.5cm}
 \epsfxsize=7.5cm
 \epsfysize=7.5cm
 \epsffile{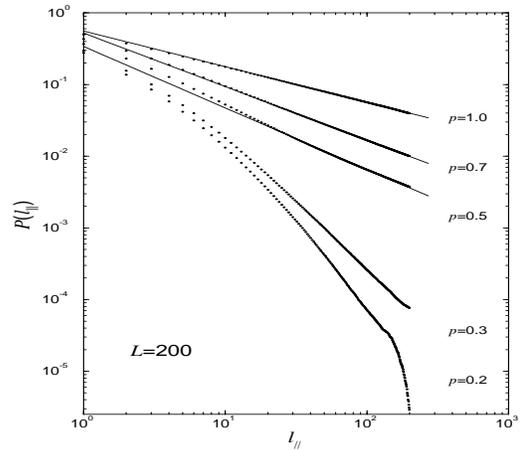} 
 \caption{Double logarithmic plot of the probability distribution 
	  $P(l_\parallel)$ of avalanches of length $l_\parallel$
	  for various values of $p$ as indicated. The solid lines correspond 
          to power laws.
 \label{length}}
 \end{minipage}
\end{figure}

\begin{figure}
 \begin{minipage}[t]{7.5cm}
 \epsfxsize=7.5cm
 \epsfysize=7.5cm
 \epsffile{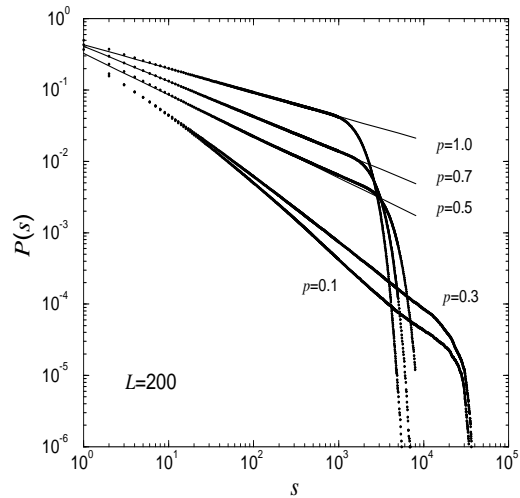} 
 \caption{Double logarithmic plot of the probability distribution 
	  $P(s)$ of avalanches of size $s$
	  for various values of $p$ as indicated. The solid lines correspond 
          to power laws.
 \label{size}}
 \end{minipage}
\end{figure}

In this section we examine how the SOC 
behavior of the Dhar model vanishes with decreasing $p$.
In the critical state, due to the 
absence of a characteristic length scale, the probability $P(l_{\parallel})$ for 
an avalanche longer than a certain length ${l_{\parallel}}$ 
obeys a power-law, i.e., 
\FL
\begin{equation}
P({l_{\parallel}}) \sim {l_{\parallel}}^{-\kappa} . 
\label{eq:length_prob}
\end{equation} 
Similarly, the probability distribution of an avalanche size larger than 
$s$ exhibits a power law with another exponent via
\FL
\begin{equation}
  P(s) \sim s^{-\tau} , 
\label{eq:size_prob}
\end{equation} 
where $s$ is the number of different sites that relaxed in one event.
Multiply relaxed sites are counted only once in this distribution.
Double logarithmic plots of the distributions $P({l_{\parallel}})$
and $P(s)$ for various values of the 
parameter $p$ and $L=200$ are shown in Fig.~\ref{length} and Fig.~\ref{size}.

\begin{table}
\vspace{0.1cm}
\caption{The exponents for different values of toppling parameter $p$.
	 Due to a lack of a power law behavior of the corresponding
	 quantities the exponents $\tau, \kappa$ and $\theta$
	 are not defined for $p<0.5$.}
\label{table_exp}
\begin{tabular}{cccccccc}
$p$ & $d_{||}$ & $\zeta$  &$z$ & $\tau$ & $\kappa$ & $\theta$ \\  
\tableline
1.0  & 1.47  & 0.48 &  1.00  &  0.34 &  0.50   &  0.51  \\
0.9  & 1.49  & 0.53 &  1.03  &  0.46 &  0.68   &  0.61  \\
0.8  & 1.51  & 0.55 &  1.06  &  0.47 &  0.72   &  0.65  \\ 
0.7  & 1.53  & 0.56 &  1.10  &  0.49 &  0.75   &  0.67  \\
0.6  & 1.56  & 0.58 &  1.15  &  0.53 &  0.79   &  0.69  \\ 
0.5  & 1.60  & 0.62 &  1.21  &  0.58 &  0.86   &  0.75  \\
0.4  & 1.68  & 0.69 &  1.33  &  -    &  -      &  -     \\
0.3  & 2.01  & 1.02 &  1.70  &  -    &  -      &  -     \\
0.2  & 2.11  & 1.14 &  1.77  &  -    &  -      &  -     \\
0.1  & 2.08  & 1.18 &  1.76  &  -    &  -      &  -     \\
\end{tabular}
\end{table}

One can see from Fig.~\ref{length} that the
probability distribution $P(l_\parallel)$ displays no
cut-off in the critical height regime.
For $p<p_c$ a sharp cut-off length occurs in the distribution
$P(l_\parallel)$.
The dynamics is now dominated by the toppling processes due to the
critical slope condition and thus the edge of the system
($i=L$) influences the topplings at a certain number of rows
close to the edge.
This behavior is typical for critical slope models.

\begin{figure}
 \begin{minipage}[t]{7.5cm}
 \epsfxsize=7.5cm
 \epsfysize=7.5cm
 \epsffile{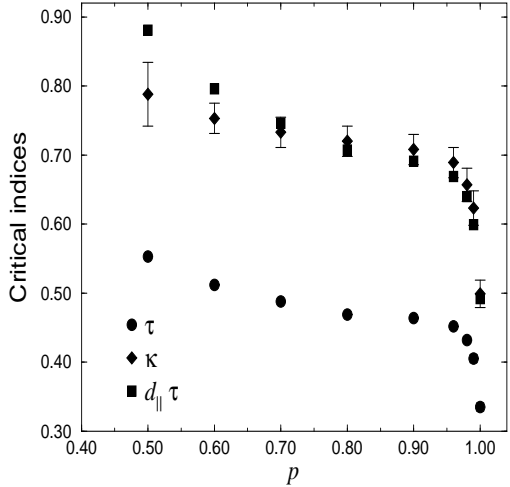} 
 \caption{Critical exponents $\kappa$ and $\tau$ determined from
	  the power law fits according to Eq.~\ref{eq:length_prob} 
	  and Eq.~\ref{eq:size_prob}.
	  In the case of the exponent $\tau$ the error bars are smaller
	  than the symbols. The scaling equation Eq.~\ref{eq:scal_rel} holds
	  for $p>0.6$. 
 \label{scal_rel}}
 \end{minipage}
\end{figure}

Power law behavior according to Eq.~\ref{eq:length_prob} 
and Eq.~\ref{eq:size_prob} occurs only in the
critical height regime.
The critical exponents $\kappa$ and $\tau$ are determined
by the slopes of the linear sections of these curves for
$p\geq0.5$.
For lower values of $p$ we found a finite curvature  in the
log-log plots of the distributions, indicating that the
power-law behavior is lost for a range of
values of $p$ preceding the transition point $p_c$.
The obtained values of the exponents for certain values of
$p$ are shown in Fig.~\ref{scal_rel} and are listed 
in Table~\ref{table_exp}.
In the limit $p=1$ the numerical values of the exponents
coincide with the exact values \cite{DHAR} $\kappa=\frac{1}{2}$
and $\tau=\frac{1}{3}$.

We should emphasize that the above analysis is not a sufficient
evidence that the model exhibits SOC for $p\geq0.5$.
To improve the accuracy of the results we analyse
the effects due to the finite size of the systems
by using simple finite-size scaling \cite{KADANOFF}:
\FL
\begin{equation}
P(s,L) = L^{- {\beta_s}} \, g( \, s \, L^{- {\nu_s}} \,). 
\label{fss}
\end{equation}
In order that Eq.~(\ref{eq:size_prob}) is recovered for large
$L$ the universal function $g(x)$ must obey a power-law
behavior for $x\to 0$ and the three 
exponents $\beta_s$, $\nu_s$
and $\tau$ fulfill the relation 
\FL
\begin{equation}
{\beta_s} = {\nu_s} \, \tau.
\label{sr_1}
\end{equation}
The scaling exponent ${\nu_s}$ can be derived from the fractal dimension $d_{\parallel}$:
For each value of $p$ we find a cut-off of the distribution 
taking place at a size $s_{max}$ which depends on $L$.
If finite-size scaling works all distributions $P(s,L)$, including their
cut-offs, have to collapse, i.e. the argument
of the universal function $g$ has to be constant:
\FL
\begin{equation}
s_{max} \, L^{-{\nu_s}} \, = \, \mbox{const}.
\label{s_max}
\end{equation}
On the other hand we know from Eq.~(\ref{eq:size_length}) that 
$\langle s \rangle({l_{\parallel}})$ 
scales with the length ${l_{\parallel}}$.
Thus $\langle s_{max} \rangle$ scales with $L$ in the same way and both 
exponents are identical.
In this way we have calculated the scaling exponents 
\FL
\begin{equation}
{\beta_s} =  d_{\parallel} \, \tau \hspace{1cm} \mbox{and} \hspace{1cm} {\nu_s} = 
d_{\parallel}
\label{eq:fss_expo}
\end{equation}
from the measured exponents $d_{\parallel}$ and $\tau$.

\begin{figure}
 \begin{minipage}[t]{7.5cm}
 \epsfxsize=7.5cm
 \epsfysize=7.5cm
 \epsffile{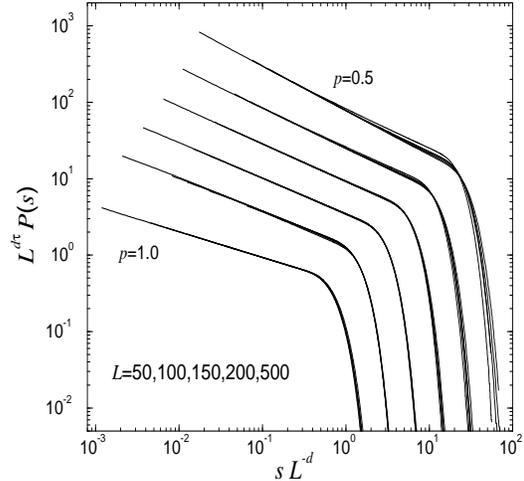} 
 \caption{Finite size scaling analysis of the avalanche size
	  distribution $P(s)$ for $p=1.0, 0.9, 0.8, 0.7, 0.6, 0.5$ and
	  five different system sizes $L$. 
	  The curves corresponding to
	  $p<1$ are shifted in the upper right direction.
	  The finite size scaling ansatz works quite well for $p>0.6$.
 \label{fss_size}} 
 \end{minipage}
 \end{figure}

In Fig.~\ref{fss_size} we show the results of the scaling plot for  
$p \geq 0.5$ and five different automata sizes. 
Finite-size scaling works well for $p \geq 0.7$.
In the case of $p\leq 0.6$ the deviations grow, i.e., the different
curves do not completely collapse.

In the following we will proof a scaling relation among
the exponents $\tau$, $\kappa$
and $d_\parallel$, indicating that the exponents
of the distributions $P(s)$ and $P({l_{\parallel}})$ are not independent.
The relation 
\FL
\begin{equation}
p(l_\parallel)  \; \mbox{d}l_\parallel \; \sim \;
p(l_\parallel)  \; \frac{\mbox{d}l_\parallel}{\mbox{d}s} \; \mbox{d}s \; \sim \;  
p(s) \; \mbox{d}s 
\end{equation}
and Eq.~(\ref{eq:size_length}) lead to the scaling relation
\begin{equation}
\kappa = d_{\parallel} \, \tau \; .  
\label{eq:scal_rel}
\end{equation}
Here
$p(l_{\parallel})$ and $p(s)$  represent the probability density 
of an avalanche of
length $l_\parallel$ and size $s$, i.e.~they obey the power-laws 
$p({l_{\parallel}}) \sim {l_{\parallel}}^{-\kappa-1}$ and 
$p(s) \sim s^{-\tau-1}$, with $\kappa$ and
$\tau$ defined in (\ref{eq:length_prob}) and (\ref{eq:size_prob}), 
respectively.

In Fig.~\ref{scal_rel} the product $d_\parallel \, \tau$ is also 
shown and compared with the values
of the exponent $\kappa$ for various values of $p$. 
We conclude that the scaling relation (\ref{eq:scal_rel}) holds 
for $p \geq 0.7$. 
These results confirm the finite-size scaling analysis and suggest
that criticality is lost at a certain point between $p=0.6$
and $p=0.7$.
We expect that preceding the transition point $p_c$ the
occurrence of toppling processes due to the critical
slope condition increases and results in a loss of criticality
which is connected to the pure critical height behavior.

We have also investigated the avalanches durations $t$ and found
that the average duration scales with the length $l_\parallel$
according to 
\FL
\begin{equation}
\langle t \rangle _{l_\parallel} \sim \l_\parallel^z.
\end{equation}
Analogous to the size $s$ the probability distribution of the
duration obeys a power-law behavior 
\FL
\begin{equation}
P(t) \sim t^{-\theta}
\end{equation}
for $p \geq 0.7$.
In the same area of $p$ finite-size scaling works quite well
and the corresponding scaling relation $\kappa=z \theta$ is
fulfilled.
The obtained values of the exponents $\theta$ and $z$  are 
listed in Table~\ref{table_exp}.
These measurements confirm our previous conclusion that the system
exhibits SOC for $p \geq 0.7$ and not in the whole critical height regime.

\section{Conclusions}

We have studied numerically a sandpile model 
with stochastic dynamics which exhibits a nonequilibrium  phase 
transition as the parameter $p$ (representing static friction) 
is varied.
The probability parameter allows us to tune the dynamics 
from a critical height to a critical slope behavior.
The average net slope $\langle \sigma \rangle$  plays the role of the 
order parameter and obeys a power law dependence in $p_c-p$.  
In contrast to the diverging penetration depth both the linear
and nonlinear susceptibility shows no singular behavior by approaching the 
transition point $p_c$.   

\begin{figure}[p]
 \begin{minipage}[t]{7.5cm}
\unitlength0.65cm
\begin{center}
\begin{picture}(12,9)

\tiny
\put(1,1){\line(1,0){10}}
\put(1,1.2){\line(0,-1){0.3}}
\put(2,1){\line(0,-1){0.1}}
\put(3,1){\line(0,-1){0.1}}
\put(4,1){\line(0,-1){0.1}}
\put(5,1){\line(0,-1){0.1}}
\put(6,1){\line(0,-1){0.1}}
\put(7,1){\line(0,-1){0.1}}
\put(8,1){\line(0,-1){0.1}}
\put(9,1){\line(0,-1){0.1}}
\put(10,1){\line(0,-1){0.1}}
\put(11,1.1){\line(0,-1){0.2}}

\put(1,0.5){\makebox(0,0)[b]{0}}
\put(3,0.5){\makebox(0,0)[b]{0.2}}
\put(5,0.5){\makebox(0,0)[b]{0.4}}
\put(7,0.5){\makebox(0,0)[b]{0.6}}
\put(9,0.5){\makebox(0,0)[b]{0.8}}
\put(11,0.5){\makebox(0,0)[b]{1}}
\put(6,0.1){\makebox(0,0)[b]{\Large{\it{p}}}}

\put(0,1){\makebox(2,1){Random Walk}}
\put(10,1.2){\makebox(2,1){Dhar and}}
\put(10,0.8){\makebox(2,1){Ramaswamy}}

\put(1,1.8){\line(0,1){4.1}}
\put(11,1.9){\line(0,1){5.0}}

\put(1,2.5){\vector(1,0){2.92}}
\put(11,2.5){\vector(-1,0){7.06}}
\put(1.5,2.1){\makebox(2,0.8)[t]{Slope regime}}
\put(6.5,2.1){\makebox(2,0.8)[t]{Height regime}}

\multiput(3.93,1)(0,0.3){8}{\line(0,1){0.2}}

\put(2.93,3){\makebox(2,2)[t]{$\sigma \sim \epsilon^{\beta}$}}
\put(2.93,2.5){\makebox(2,2)[t]{$\chi_, \; \chi_{_{nl}}$}}
\put(2.93,2){\makebox(2,2)[t]{$\lambda \sim \epsilon^{-\nu_{\parallel}}$}}
\put(2.93,3.3){\framebox(2,2){}}

\put(1.1,6){\vector(1,0){2.82}}
\put(11,6){\vector(-1,0){7.06}}
\put(1.5,5.6){\makebox(2,0.8)[t]{Isotropic}}
\put(6.5,5.6){\makebox(2,0.8)[t]{Anisotropic}}

\multiput(3.93,5.3)(0,0.3){12}{\line(0,1){0.2}}

\put(10.9,7){\vector(-1,0){6.96}}
\put(6.5,6.6){\makebox(2,0.8)[t]{Fractal}}

\put(11,8){\vector(-1,0){3.5}}
\put(8.25,7.6){\makebox(2,0.8)[t]{SOC}}
\put(6.8,7.5){\makebox(1,1){?}}

\put(1,6){\circle{0.2}}
\put(1,6.1){\line(0,1){2.7}}

\put(11,7){\circle{0.2}}
\put(11,7.1){\line(0,1){1.7}}

\end{picture}
\end{center}
 \caption{Phase diagram of the model which illustrates the major 
	  properties of the investigated system. The order parameter
	  $\sigma$ and the penetration depth $\lambda$ depend on
	  $\epsilon=|p-p_c|$.
 \label{phase_dia}} 
 \end{minipage}
\end{figure}
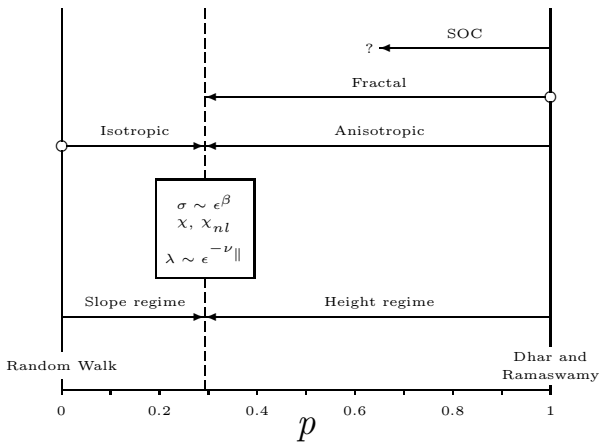

We found that the nonequilibrium phase transition is accompanied
by a transition from an anisotropic to an isotropic behavior of the
avalanches.
Fractality occurs only in the case of the 
anisotropic shape of the avalanches.
 
Finite size scaling analysis works for the probability distributions
of the avalanches and scaling relations 
hold for $p \geq 0.7$ indicating that the system exhibits SOC in this
region.
The corresponding exponents have a nonuniversal i.e. $p$-dependent
behavior.

Due to long computation time, we did not simulate lattices larger than
$L=500$ in order to check if the power-law behavior is still maintained
for values of $p<0.6$.  The problem of disappearance of the SOC on 
approaching the nonequilibrium phase transition remains for future study.

A complete illustration of the behavior of our sandpile model
is depicted in Fig.~\ref{phase_dia}.

\acknowledgments
This work was supported by the 
Bundesministerium f\"ur Bildung, Wissenschaft, Forschung
und Technologie, by the
Deutsche Forschungsgemeinschaft through
Sonderforschungsbereich 166, Germany, and by the Ministry of Science
and Technology of the Republic of Slovenia.


\vspace{-0.5cm}
\begin{references} 
\vspace{-0.9cm}
\bibitem[*]{SvenEmail} E-mail: sven@hal6000.thp.uni-duisburg.de
\bibitem[\dagger]{BosaEmail} E-mail: Bosiljka.Tadic@ijs.si
\bibitem[\diamond]{UsadelEmail} E-mail: usadel@hal6000.thp.uni-duisburg.de

\bibitem{BAK} P.~Bak, C.~Tang and K.~Wiesenfeld, Phys.~Rev.~Lett.  {\bf 59}, 381 (1987). \\   
              P.~Bak, C.~Tang and K.~Wiesenfeld, Phys.~Rev.~A {\bf 38}, 364 (1988).

\bibitem{GRINSTEIN} G.~Grinstein, in
		    {\it Scale Invariance, Interfaces and Non-Equilibrium Dynamics},
		    edited by A.~McKane {\it et al.}, 
		    NATO ASI Series~B: Physics Vol.~344, Plenum Press (1995).
\bibitem{FLYVBJERG} H.~Flyvbjerg, in 
		    {\it Scale Invariance, Interfaces and Non-Equilibrium Dynamics},
		    edited by A.~McKane {\it et al.}, 
		    NATO ASI Series~B: Physics Vol.~344, Plenum Press (1995).

\bibitem{DHAR}      D.~Dhar and R.~Ramaswamy, Phys.~Rev.~Lett. {\bf 63}, 1659 (1989).

\bibitem{MANNA}     S.~S.~Manna, Physica~A {\bf 179}, 249 (1991).

\bibitem{KADANOFF}  L.~P.~Kadanoff, S.~R.~Nagel, L.~Wu and S.~M.~Zhou, Phys.~Rev.~A  
                    {\bf 39}, 6524 (1989).

\bibitem{LUEB_1}    S.~L\"ubeck, K.~D.~Usadel, Fractals {\bf 1}, 1030 (1993).

\bibitem{LUEB_2}    S.~L\"ubeck, B.~Tadi\'c and K.~D.~Usadel, in {\it Fractal 
		    Reviews in the Natural and Applied Sciences}, 
                    Proceedings of the Fractal conference in Marseille Feb. 1995,
                    edited by M.~M.~Novak, Chapman \& Hall (1995). 

\bibitem{BINDER}    K.~Binder and J.-S.~Wang, J.~Stat.~Phys. {\bf 55}, 87 (1989).

\bibitem{KINZEL}    W.~Kinzel, Ann.~Israel~Phys.~Soc., {\bf 5}, 425 (1983).

\bibitem{KAISER}    C.~Kaiser and L.~Turban, J.~Phys.~A: Math.~Gen.~{\bf 27},
		    L579 (1994).



\end{references}
\end{document}